# Exact analytic characteristic initial data for axisymmetric, non-rotating, vacuum space-times, with an application to the binary black hole problem.

Ewald Wessels[†]
16 Alexandra Rd., Cape Town 7700, South Africa.

**Abstract**  *Bondi's approach to the construction of a coordinate system is used with a different choice of gauge, in accordance with which the radial coordinate r is an affine parameter, to cast the metric tensor into a form suitable for use with the Newman-Penrose null tetrad formalism. The choice of tetrad has the result that the equations and all the functions that appear in them are real-valued. A group classification of the Sachs equations in this gauge leads to a unique expression for the first of the five independent elements $\Psi_0$ of the Weyl spinor, and to the corresponding exact solutions for two of the metric functions on an initial null hypersurface. A proof is presented that the result for $\Psi_0$ constitutes the appropriate characteristic initial value function for all physically realistic axisymmetric, non-rotating vacuum space-times. Integration of the field equations on the axis of symmetry when an equatorial symmetry plane is also present, and on the equatorial plane itself when time rates of change can be neglected, shows that these data produce results consistent with Newton's laws and with the Schwarzschild solution in the appropriate limits. The solution on the axis of symmetry indicates that the Weyl curvature increases without limit between black holes as their separation decreases.*

## 1. Introduction

It is well known that Einstein's vacuum field equations can be solved in principle for the metric tensor of a non-rotating axisymmetric vacuum space-time provided that certain initial data are specified. The problem of finding physically interesting solutions therefore reduces to that of finding the form of the appropriate initial data. It is the purpose of this paper to demonstrate that by combining Bondi's approach to the construction of a coordinate system [1],[2] with the Newman-Penrose (hereafter referred to as NP) spinor formulation of the field equations [3] it is possible to cast the equations into a particularly transparent form. With the use of a Lie group analysis an initial value function can be obtained from two of these equations. It will be shown (in appendix A) that this function provides the appropriate initial data for all physically realistic vacuum space-times that possess an axis of symmetry and are non-rotating (i.e. that have reflection symmetry with respect to rotation about the axis). One form of the initial value function yields the Weyl tensor that is implied by Newton's laws as an approximation on the axis of symmetry. It yields a pair of Schwarzschild solutions, and a second form yields a single Schwarzschild solution, as an approximation in the appropriate case.

## 2. Coordinate system and tetrad structure.

Bondi's coordinates are based on a set of null cones generated from the timelike curve followed by a point O that remains on the axis of symmetry. The null directions at O are labelled with spherical coordinates $\theta$ and $\phi$. Radial distances are measured by the parameter r that characterises individual generators of the null cone, and the coordinate set is completed with a retarded time v, which is constant on each null hypersurface. Assign numerical references 1,2,3,4 to v, r, $\theta$, and $\phi$ in that order. "Geometrised" units of measure, in which c = G = 1, and a space-time signature of -2, will be employed throughout.

Null geodesics are characterised by v, $\theta$, $\phi$ = const. It follows from the fact that only r varies along an outgoing null geodesic that $g_{22}$ must be zero. Symmetry under the reflection $\phi \rightarrow -\phi$ requires that $g_{14} = g_{24} = g_{34} = 0$.

The absolute derivative of the tangent to a curve, along the same curve, reduces to the following expression if the curve parameter r is used as one of the coordinates:

$$\ell^\mu \nabla_\mu \ell^\nu = \Gamma^\nu{}_{22} \qquad (2.1)$$

---

[†] e-mail address: wessels@iafrica.com

It follows that if the curve is a geodesic $\Gamma^1_{22} = \Gamma^3_{22} = 0$. Bondi et. al. [2] demonstrate that this implies that the ratio $g_{23}/g_{12}$ must be independent of r. This allows $g_{23}$ to be eliminated by the transformation:

$$du = \lambda \left( dv + \frac{g_{23}}{g_{12}} d\theta \right) \qquad (2.2)$$

where $\lambda$ is a suitable function of v and $\theta$.

Bondi et. al. use the remaining coordinate freedom to make r an area distance. If the coordinate freedom is used, instead, to transform r in such a way that $g_{12} = 1$, the consequence is that $\Gamma^2_{22} = 0$ and r becomes an affine parameter. The metric tensor now has the canonical form used by Newman, Penrose and their co-workers. These coordinates look like Robinson-Trautman [4] coordinates but there is a significant difference in the fact that the current coordinates are related through (2.2) to a set of null cones that have their vertices on the axis of symmetry, whereas Robinson-Trautman coordinates do not require the existence of null cones. As pointed out by Penrose [5], the presence of the vertex on a null cone provides a common reference point to which field variables may be referred by parallel transport along the generators of the null cone.

If it is assumed that the manifold has a spinor structure [6], a null tetrad may now be introduced at every point of the space-time manifold, using as reference the tangent vector $\ell^\mu$ to the null geodesics that form the coordinate rays. A spinor structure requires that the manifold should have a unique orientation at every point. To achieve this, the neighbourhood of the origin must be simply connected and may not be extended beyond any points at which the coordinate rays form cusps (i.e. where the null geodesics cross one another). This means that in general the space-time will have disjoint coordinate patches.

The nature of the coordinates and the form of the metric tensor established above means that:

$$\ell^\mu = \delta^\mu_2, \qquad \ell_\mu = \delta^1_\mu \qquad (2.3)$$

where $\delta$ is the Kronecker delta.

The tetrad is completed with a second real null vector $n^\mu$ normalised so that $n^\mu \ell_\mu = 1$ and a complex null vector $m^\mu$ together with its complex conjugate $\overline{m}^\mu$ (for the details see NP [3] ). The tetrad vectors are related to the elements of the metric tensor by the relation

$$g^{\mu\nu} = \ell^\mu n^\nu + n^\mu \ell^\nu - m^\mu \overline{m}^\nu - \overline{m}^\mu m^\nu \qquad (2.4)$$

A similar relation holds for the components of the covariant form of the metric tensor.

To satisfy the orthogonality relations the tetrad vectors must have the following form:

$$n^\mu = \delta^\mu_1 + U\delta^\mu_2 + X^i \delta^\mu_i \qquad (i = 3,4) \qquad (2.5)$$

and

$$m^\mu = \omega \delta^\mu_2 + \xi^i \delta^\mu_i \qquad (2.6)$$

where, using the notation of NP and their co-workers, U, $X^i$, $\omega$, and $\xi^i$ are unknown, complex-valued functions of u, r, and $\theta$. Using (2.4) together with (2.5) and (2.6) gives the elements of the metric tensor in terms of these functions.

The vector $n^\mu$ can be chosen so that at the origin its spacelike projection lies in a plane containing the axis of symmetry. Symmetry then requires that it must remain in this plane as the tetrad propagates outwards. One can therefore put $X^4 \equiv 0$ so that there is no ambiguity in putting $X^3 \equiv X$. The complex null vector $m^\mu$ is defined relative to a pair of real vectors which span the spacelike hyperplane tangent to the surfaces u = const. If the imaginary axis is chosen to lie along the direction of the coordinate $x^4$ then $\xi^3$ becomes a real-valued function. It is easy to show that all the metric functions are then real-valued except for $\xi^4$ which is pure imaginary.

For later convenience put $\xi^3 = 1/q$ and $\xi^4 = i/Q$ where both q and Q are real-valued functions of the coordinates. The metric tensor resulting from this analysis is:

$$ds^2 = (-2U - \tfrac{1}{2} X^2 q^2 + 2X\omega q)\, du^2 + 2\, du\, dr - (2\omega q - q^2 X)\, du\, d\theta - \tfrac{1}{2} q^2\, d\theta^2 - \tfrac{1}{2} Q^2\, d\phi^2$$

$$g^{\mu\nu} = \begin{pmatrix} 0 & 1 & 0 & 0 \\ 1 & 2(U - \omega^2) & X - \dfrac{2\omega}{q} & 0 \\ 0 & X - \dfrac{2\omega}{q} & \dfrac{-2}{q^2} & 0 \\ 0 & 0 & 0 & \dfrac{-2}{Q^2} \end{pmatrix} \qquad (2.7)$$

As is the case with Bondi's metric tensor this metric contains essentially four unknown functions. The apparent redundancy following from the fact that five unknown functions survive from the specification of the tetrad vectors will be seen later to disappear when the condition for parallel propagation of the tetrad is imposed.

It is easy to see that the functions $\omega$ and $X$ measure the focusing of the coordinate rays. For an asymptotically flat space-time they must tend to zero at infinity. When it can be assumed that the metric tensor will approach the Schwarzschild form as $r \to \infty$, then the affine parameter $r$ must at the same time tend to the Schwarzschild radial coordinate (which is an affine parameter for radial null geodesics in the Schwarzschild space-time). Also, in this case one must have

$$U \to -\tfrac{1}{2} + \tfrac{m}{r}, \qquad q \to \sqrt{2}\, r, \qquad Q \to \sqrt{2}\, r \sin\theta \qquad \text{as } r \to \infty \qquad (2.8)$$

where m tends to the Schwarzschild mass of the system.

### 3. The Newman-Penrose equations

Expressed in terms of the metric tensor (2.7), all the spin coefficients, all the elements of the Weyl spinor, and all the NP equations are real-valued. This result greatly simplifies the equations without any loss of transparency. The eight "metric equations" that relate the spin coefficients to the elements of the metric tensor are all satisfied identically. To avoid repeating the numerous NP equations themselves they will be referred to by the equation numbers employed in [3].

The condition for parallel propagation of the tetrad is that the spin coefficient $\pi$ must vanish. This provides the relation:

$$\frac{q}{2}\frac{\partial X}{\partial r} + \frac{\omega}{q}\frac{\partial q}{\partial r} + \frac{\partial \omega}{\partial r} = 0 \qquad (3.1)$$

The NP equations (6.11a) to (6.11r) are reduced to 12 equations when (3.1) is taken into account. It can be demonstrated that any solution of these equations, in the current coordinate system, will be regular on the axis of symmetry. This demonstrates consistency with the assumption of regularity inherent in their derivation and that of the metric tensor (2.7). The analysis also produces a number of results that are required in section 5.

The integrability of the equations on the axis of symmetry follows trivially from Frobenius's theorem [7]. (Note that, while there is a coordinate singularity on the axis, points on the axis can be included in the u,r,$\theta$ hyperplane by analytic continuation.). The behaviour of the metric tensor and the tetrad vectors on opposite sides of the axis of symmetry shows that the functions $\omega$, X, and Q must be odd under the reflection $\theta \to -\theta$. The definitions of the elements of the Weyl spinor show that the same must be true of the functions $\Psi_1$ and $\Psi_3$. If these functions are analytic then the functions themselves and all their even partial derivatives with respect to $\theta$ must vanish on the axis. The opposite is true of the functions q, U, $\Psi_0$, $\Psi_2$, and $\Psi_4$. These functions must be even so that all their odd partial derivatives with respect to $\theta$ must vanish on the axis.

The spin coefficients $\rho$ and $\sigma$, also called the "optical scalars", provide additional information. In the current coordinates the definitions of these functions lead to the following expressions:

$$\rho + \sigma = -\frac{1}{q}\frac{\partial q}{\partial r} \qquad (3.2)$$

$$\rho - \sigma = -\frac{1}{Q}\frac{\partial Q}{\partial r} \qquad (3.3)$$

Sachs [8] first showed that $\sigma$ measures the shear of the null congruence $\ell^\mu$ while $\rho$ measures the expansion of the congruence. On the axis of symmetry $\sigma$ must be zero.

Equations (6.11a) and (6.11b) in NP were discovered by Sachs [8]. They read:

$$\frac{\partial \rho}{\partial r} = \rho^2 + \sigma^2 \qquad (3.4)$$

$$\frac{\partial \sigma}{\partial r} = 2\rho\sigma + \Psi_0 \qquad (3.5)$$

It follows immediately that on the axis of symmetry

$$\Psi_0 = 0 \quad \text{and} \quad \rho = \frac{1}{\rho^0 - r} \qquad (3.6)$$

where $\rho^0$ is a function that is independent of r. The first of these two results also follows directly from the Petrov classification of the space-time on the axis of symmetry[†]. The possibility that $\rho = 0$ everywhere can be excluded immediately on the basis of elementary flatness, since a null cone in flat space is characterised by $\rho \neq 0$.

Subtracting (6.11d) from (6.11e) in NP, multiplying by Q and then allowing the functions to tend to their values on the axis of symmetry gives a first order ordinary differential equation for $\partial Q/\partial\theta$ (regarded as a function of u and r defined on the axis of symmetry). Without loss of generality the analysis can be confined to the first quadrant so that the value of $\theta$ on the axis is single-valued and constant, with the value $\theta = 0$. Solving the equation then gives the result that $\partial Q/\partial\theta = f(u)\,q$ where f(u) is a function of integration. Multiplying (6.11m) by Q and letting the functions tend to their values on the axis gives $\partial f/\partial u = 0$. It is possible therefore to adopt f(u) = 1 as an initial value, so that for all u on the axis of symmetry:

$$\frac{\partial Q}{\partial \theta} = q \qquad (3.7)$$

This is the condition for regularity on the axis of symmetry. Its validity makes it possible to use l'Hôpital's rule to evaluate the indeterminate forms 0/0 that appear in the NP equations when Q is in the denominator and the equations are evaluated on the axis in the current coordinates.

## 4. Null hypersurface initial data

The NP equations can be divided into 'hypersurface equations' that involve no derivatives with respect to u, and the remaining equations that do involve u derivatives. The equations can be integrated by starting from an initial function, specified on a single null hypersurface. The values of all the metric functions on the initial hypersurface can be determined from the hypersurface equations. The first derivatives with respect to u can then be determined from the remaining equations. Higher u derivatives can be determined by differentiating all the equations with respect to u and repeating the cycle. This allows a formal expansion in powers of u to be generated for each of the metric functions. The conditions under which such expansions converge have been investigated by Friedrich [14].

In the current formulation of the field equations the natural candidate for an initial function is $\Psi_0$, the first of the five independent elements of the Weyl spinor. This follows from an analysis carried out by Penrose [5] who demonstrated that the values of $\Psi_0$ on the points of an initial null hypersurface in an analytic empty space-time, together with Einstein's field equations, determine the entire field. Penrose also demonstrated that these initial data can be chosen freely without any constraints (when there are no constraints on the geometry of the space-time), except perhaps that they should be analytic.

---

[†] The author is grateful to Malcolm MacCallum for pointing out this result

The hypersurface equations that provide the first step in the integration of the field equations in the current form are the Sachs equations (3.4) and (3.5). In the coordinates used here they take a particularly simple form:

$$\frac{\partial^2 q}{\partial r^2} + \Psi_0 q = 0 \quad \text{and} \quad \frac{\partial^2 Q}{\partial r^2} - \Psi_0 Q = 0 \tag{4.1}$$

Since $\Psi_0$ is a function of r and $\theta$ that is not determined by the NP equations, it can be specified independently of these equations on the initial null hypersurface. Also, the two Sachs equations involve differentiation with respect to r only. They can therefore be treated as a closed system of ordinary differential equations that determine the r-dependence of q and Q once the character of the space-time has been prescribed by the choice of $\Psi_0$ on the initial hypersurface. Much attention has been directed at determining the nature of asymptotic expansions that constitute physically realistic choices for the form of the initial data, mainly with a view to exploring the nature of gravitational radiation at infinity. NP and their co-workers chose an expansion in terms of powers of 1/r with the condition that $\Psi_0 = O(r^{-5})$ for a general (not necessarily axially symmetric) radiation metric. More recently it has been argued [9] that polyhomogeneous space-times, admitting expansions in terms of $r^{-j} \log^i r$, may be more appropriate.

As an alternative to prescribing a series expansion of $\Psi_0$ it seems natural to carry out a group classification [10] of the Sachs equations (4.1) to determine whether there are any functional forms for $\Psi_0$ that would extend the principal Lie algebra possessed by the pair of equations. (The principal Lie algebra of a set of differential equations is the Lie algebra possessed by the equations regardless of the form of an undetermined function.) On the initial null hypersurface the value of $\Psi_0$ is independent of q, Q, and their derivatives [5]. It is therefore possible to put $\Psi_0 = \Psi_0(r,\theta)$ in the space of differential functions [10] for these equations. It follows that the principal Lie algebra of (4.1) corresponds to the pair of scale transformations of q and Q. This is consistent with the fact that Einstein's field equations are invariant under scale transformations of the elements of the metric tensor [11]. (However, note that this holds for the full set of field equations only when the same re-scaling factor is applied to all the elements of the metric tensor.) It turns out that if $\Psi_0$ is regarded as invariant, additional symmetries appear if, and only if, $\Psi_0$ takes the following form [12]. The value u = 0 designates the single null hypersurface on which the result applies.

$$\Psi_0 \big|_{u=0} = \frac{a_0}{\left(b_2 r^2 + b_1 r + b_0\right)^2} \tag{4.2}$$

where $a_0$ and the $b_i$ are functions that are independent of r.

The equations then possess a three-dimensional Lie algebra spanned by the operators

$$X_1 = (b_2 r^2 + b_1 r + b_0)\frac{\partial}{\partial r} + rq\frac{\partial}{\partial q} + rQ\frac{\partial}{\partial Q}$$

$$X_2 = q\frac{\partial}{\partial q} \qquad X_3 = Q\frac{\partial}{\partial Q} \tag{4.3}$$

where $X_2$ and $X_3$ constitute the principal Lie algebra.

The analysis in appendix A demonstrates that the hypersurface equations for any solution of the field equations representing a physically realistic non-rotating axisymmetric space-time must be invariant under a Lie point transformation that transforms the coordinate r as the operator $X_1$ does, and that leaves the function $\Psi_0$ invariant. The only functional form of $\Psi_0$ that allows (4.1) to admit such a symmetry is (4.2). It follows that $\Psi_0$ must take the form (4.2) on an initial null hypersurface for any solution representing a physically realistic non-rotating axisymmetric vacuum space-time. The analysis in appendix A also indicates that the $b_i$ must be constants. The function $a_0$ arises as a function of integration and is, in general, a function of $\theta$.

Penrose's result [5] concerning initial null data is extended by a theorem due to Friedrich [14] that proves the existence of analytic solutions of Einstein's field equations for analytic initial data on past null infinity and on an incoming null hypersurface. (Note that the vertex corresponding to any particular null cone is a point that is, strictly speaking, not part of the null hypersurface associated with that cone. Regularity conditions therefore have to be imposed at the vertex to reflect, by analytic continuation, the regularity that is assumed to exist on the null hypersurface associated with the past null cone.) It follows that there are exact analytic solutions of Einstein's field equations corresponding to the initial null data provided by (4.2). These solutions can be found, in principle, by the procedure outlined above, but in practice the complete integration of the equations remains a formidable computational problem.

The solutions of (4.1) with $\Psi_0$ in the form (4.2) take five different forms depending on the relative magnitudes of $a_0$ and the $b_i$. Assuming $b_2 \neq 0$ then it is possible to set $b_2 = 1$. If $b_1^2 - 4b_0 + 4|a_0| < 0$ the solution can be put in the form:

$$q = q^0 \left(r^2 + b_1 r + b_0\right)^{\frac{1}{2}} \cos\left[\left\{\frac{4(b_0 + a_0) - b_1^2}{4b_0 - b_1^2}\right\}^{\frac{1}{2}} \arctan\left(\frac{\sqrt{4b_0 - b_1^2}}{2r + b_1}\right) + \phi_q\right] \qquad (4.4)$$

where $q^0$ and $\phi_q$ are functions of integration that are independent of r.

The corresponding solution for Q is similar except that the sign of $a_0$ is reversed. For the case $b_1^2 - 4b_0 b_2 - 4|a_0| > 0$ the polynomial in the denominator of (4.2) has two unequal roots at which $\Psi_0$ has singularities. Putting $b_2 = -1$ (so that the expression under the square root sign remains positive in the application to be discussed below), the solution in this case can be put in the form:

$$q = q^0 \left(b_0 + b_1 r - r^2\right)^{\frac{1}{2}} \sinh\left(\left(\frac{1}{4} - \frac{a_0}{b_1^2 + 4b_0}\right)^{\frac{1}{2}} \ln\left\{\frac{\left[\left(b_1^2 + 4b_0\right)^{\frac{1}{2}} + b_1 - 2r\right]^2}{4(b_0 + b_1 r - r^2)}\right\} + \phi_q\right) \qquad (4.5)$$

The corresponding solution for Q is similar except that the sign of $a_0$ is reversed. Other forms of the solution are given in appendix B.

When the coordinate r can extend to infinity without encountering singularities, as is the case when the solution takes the form (4.4), then q and Q tend to their flat-space values as r tends to infinity while $a_0$ and the $b_i$ remain finite (assuming appropriate values for the functions of integration). These are also the values q and Q take for the Schwarzschild solution. It is easy to show that the Schwarzschild metric is indeed a trivial solution of the equations corresponding to the case $a_0 = 0$.

### 5. The exact solution on the axis in the initial null hypersurface when $b_1 = 0$ and there is reflection symmetry about the origin with respect to r.

The transformations corresponding to (4.3) are one-parameter diffeomorphisms. Accordingly, they describe coordinate transformations within a single coordinate patch. In the case of a pair of black holes moving along the axis of symmetry, focusing of the coordinate rays would give rise to a coordinate patch that encloses an interval of the axis itself between the pair but does not extend to infinity. The solution on that part of the axis enclosed between the two black holes would therefore only describe the gravitational field measured by observers who are themselves on that part of the axis.

When the origin lies between a pair of black holes, there will be singularities on the axis on both sides of the origin. It is natural therefore to use (4.5) for this case. It follows from (4.2) and (3.6) that $a_0$ is zero on the axis of symmetry. When $a_0$ tends to zero, (4.5) can be brought into the form $q = \sqrt{2}\,r$ by assuming that $b_1 = 0$, that $\phi_q$ tends to zero as $\theta \to 0$, and that $q^0$ tends to the value $\sqrt{2}$. It follows from (3.6) and (3.2) that this value of q on the axis is appropriate to an observer who coincides with the point O from which the null cones that define the coordinate system are generated

With q independent of u on the axis of symmetry, the NP equations can be solved for $\partial X/\partial \theta$ and $\partial \omega/\partial \theta$ on the axis using NP equations (6.11f), (6.11g) + (6.11h) and (3.1) above. The resulting expressions can be used in (6.11l) (after making use of l'Hôpital's rule) combined with (6.11f) to obtain a second-order linear differential equation for U. The solution on the axis of symmetry is:

$$U\big|_{u,\theta=0} = \frac{1}{3\sqrt{2}\,r^2} \int \left(\frac{\partial^3 Q}{\partial \theta^3} - \frac{\partial^2 q}{\partial \theta^2}\right) dr - \frac{r}{3\sqrt{2}} \int \frac{1}{r^3}\left(\frac{\partial^3 Q}{\partial \theta^3} - \frac{\partial^2 q}{\partial \theta^2}\right) dr + \frac{k_1}{r^2} + \frac{k_2}{r} + k_3 r \qquad (5.1)$$

where the $k_i$ are constants of integration. The integrands can be evaluated using (4.5), taking account of the symmetry properties, discussed in section 3, of Q and q and their derivatives with respect to $\theta$ and of (3.7).

Assume $b_0$ is non-negative so that one can put $b_0 = d^2$. The constants of integration $k_1$ and $k_2$ can be evaluated using l'Hôpital's rule and the condition that U must have a finite limit as $r \to 0$. The function $\lambda$ in (2.2) can be normalised so that it is unity on the axis. It then follows from the construction of the coordinate system that $U = -1/2$ at $r = 0$. Using this condition it follows that:

$$\frac{\partial^3 Q^0}{\partial \theta^3} - \frac{\partial^2 q^0}{\partial \theta^2} = \frac{2\sqrt{2}}{d^2}\frac{\partial^2 a_0}{\partial \theta^2} - \sqrt{2} \qquad (5.2)$$

If one assumes reflection symmetry with respect to r about the origin one can require that $\partial U/\partial r \to 0$ as $r \to 0$. This condition allows $k_3$ to be determined.

From (6.11f) in NP, and the fact that $\omega$ must be zero on the axis of symmetry, it follows that:

$$\Psi_2\big|_{\theta=0} = -\frac{1}{2}\frac{\partial^2 U}{\partial r^2} \qquad (5.3)$$

Using the above results in (5.1) and then substituting in (5.3) gives an expression for $\Psi_2$ that is well-behaved as $r \to 0$ and gives the limit:

$$\lim_{r \to 0} \Psi_2\big|_{\theta=0} = \frac{1}{6\,d}\frac{\partial^2 a_0}{\partial \theta^2}\frac{1}{d^3} \qquad (5.4)$$

This expression can be compared with the Newtonian equivalent for a pair of identical massive bodies, each of mass m, located at positions +s and -s relative to an origin at their centre of mass. This can be calculated from the geodesic deviation equation applied to a pair of test particles separated by an infinitesimal spacelike displacement along the axis of symmetry and initially at rest at a distance r from the centre of mass. Taking into account that in an empty space-time the Weyl tensor equals the Riemann tensor, and that from the definition of $\Psi_2$ in the tetrad employed here $\Psi_2 = -\frac{1}{2}C_{1212}$ on the axis of symmetry, one finds, since the four-velocities of the test particles can be considered to be orthogonal to the axis of symmetry at $r = 0$:

$$\lim_{r \to 0} \Psi_2\big|_{\theta=0}(\text{Newton}) = -m\left\{\frac{1}{(s+r)^3} + \frac{1}{(s-r)^3}\right\} \qquad (5.5)$$

If d is identified with s and one puts:

$$\frac{\partial^2 a_0}{\partial \theta^2} = -12md \qquad (5.6)$$

then the limit (5.4) is identical to that of the Newtonian expression (5.5), at $r = 0$, on the null hypersurface under consideration. This is a satisfying result, considering that $a_0$ must be zero when $m = 0$ and when $d = 0$. Note also from (5.2) that, with this result, the $\theta$ derivatives of $q^0$ and $Q^0$ tend to their flat space values as $m/d \to 0$.

Substituting for the $k_i$ in (5.1) and using the identification (5.6) yields the metric function U on the axis of symmetry in a null hypersurface for the maximally symmetric case dealt with here, i.e. that of a pair of identical non-rotating black holes located on an axis of symmetry with the origin at their centre of mass:

$$U\big|_{u,\theta=0} = \frac{2m}{r\,d^2}\left(r^2 - 3d^2\right)\ln\left(\frac{d+r}{d-r}\right) - \frac{4\,m\,d}{r^2}\ln\{(d+r)(d-r)\} + \frac{8m\,d\ln(d)}{r^2} + \frac{8m}{d} - \frac{1}{2} \qquad (5.7)$$

The value of U corresponding to the Newtonian approximation (5.5) for $\Psi_2$ can be obtained by using (5.3). Choosing the constants of integration appropriately, this yields:

$$U\big|_{\theta=0}(\text{Newton}) = \frac{m}{d+r} + \frac{m}{d-r} - \frac{2m}{d} - \frac{1}{2} \qquad (5.8)$$

This clearly approximates to a pair of Schwarzschild solutions, each of mass m, centred at $r = \pm d$, for d large compared with m and $d - r$ small compared with $d + r$. Taylor expansions about the value $r = 0$ show that (5.7) and (5.8) are identical for small r/d, differing only from the fourth order terms onwards. This means that the exact relativistic metric function (5.7) would be indistinguishable from (5.8) in a sufficiently small neighbourhood of the origin.

Note that (5.7) tends to the value $8m(1 - \ln 2)/d - \frac{1}{2}$ as $r \rightarrow 0$. The fact that U does not become zero for $r \leq d$ until $d \approx 4.9\,m$ can be interpreted as follows. For a coordinate ray at an infinitesimal but non-zero angle θ the metric function q is given by (4.5). With $b_1$ zero and $b_0 = d^2$, q will have two zeros at positions $r = \pm d$. This means that the null geodesic congruence emanating from the origin has cusps at these points. The fact that U, as given by (5.7), is greater than zero until relatively small values of d can be interpreted to mean that for larger separations of the pair of black holes a cusp is encountered along the axis before a point of infinite red shift is reached.

Substituting from (5.6) in (5.4):

$$\lim_{r \rightarrow 0} \Psi_2 \Big|_{\theta=0} = \frac{-2m}{d^3} \qquad (5.9)$$

This expression is identical to the Newtonian expression for a pair of point masses each of mass m, separated by a distance 2d. However, note that in the relativistic case 2d is measured in terms of an affine parameter, and is the distance that separates the pair of points on a null geodesic along the axis of symmetry that are conjugate to the origin. An interesting conclusion follows from this result, since the null hypersurface on which the result holds is arbitrary: the "tidal force" component of the Weyl tensor on the axis of symmetry between a pair of non-rotating black holes in head-on collision increases without limit as the points on the axis at which there are cusps in the null congruence emanating from the origin approach one another. This result is probably valid more generally than for the maximally symmetric case considered here

## 6. The solution on the equatorial plane when there is reflection symmetry about the origin and time rates of change can be neglected.

The symmetry relations on the equatorial plane, which is a plane of reflection symmetry in the case being considered, are:

$$X = \omega = \frac{\partial q}{\partial \theta} = \frac{\partial Q}{\partial \theta} = \frac{\partial U}{\partial \theta} = 0 \qquad (6.1)$$

A second order differential equation for U in terms of q and Q can be obtained from NP equations (6.11g) – (6.11h) by substituting for $\Psi_2$ from (6.11f) and by using the symmetry relations in (6.1). Assuming time rates of change can be neglected the solution is exactly Schwarzschild when q and Q take the forms (2.8). Using the non-singular forms (4.4) for q and Q the solution is:

$$U\Big|_{\perp} = c_1 \frac{q}{Q} - \frac{c_2 q}{Q(4a_0 + 4b_0 - b_1^2)^{\frac{1}{2}}} \tan\left\{\left(1 + \frac{4a_0}{4b_0 - b_1^2}\right)^{\frac{1}{2}} \arctan\left(\frac{(4b_0 - b_1^2)^{\frac{1}{2}}}{b_1 + 2r}\right)\right\} \qquad (6.2)$$

where $c_1$ and $c_2$ are constants of integration and the symbol $\perp$ indicates that the equation applies on the equatorial plane in the initial null hypersurface. This expression can be expanded as a Taylor series in 1/r around the point 1/r = 0. Choosing $c_1 = -\frac{1}{2}$ and $c_2$ = -4m the result is:

$$U\Big|_{\perp} = -\frac{1}{2} + \frac{2m}{r} + \frac{a_0 - 2mb_1}{r^2} - \frac{8ma_0 + 3b_1 a_0 - 3mb_1^2}{r^3} + \cdots \qquad (6.3)$$

To first order this is a single Schwarzschild solution with mass 2m when the functions that are independent of r are finite.

## 7. Conclusion

Although the complete solution has been computed only in the initial null hypersurface for the simplest cases of the general axisymmetric problem, the results clearly show that the characteristic initial value function (4.2) produces expressions that are compatible in the appropriate limits with Newton's laws and with the Schwarzschild solution. To progress beyond the simplest cases, substantial computing power is required and it is necessary to evaluate difficult quadratures.

The prospect of detectors capable of measuring gravitational radiation from astrophysical events has led to much work on numerical solutions of Einstein's equations [15]. This work is ultimately directed at understanding the gravitational radiation expected to be produced by a pair of black holes spiralling in towards one another. Work on vacuum solutions completed to date has included the case of a pair of non-rotating black holes in head-on collision, starting from initial data that represent "two nonrotating black holes poised near one another in a momentarily stationary configuration" [16]. These initial data are based on an exact solution on a hypersurface of time symmetry, obtained by Misner [17]. The solution to the initial value problem reported here relaxes the requirement of time symmetry.

Some workers have used approximate initial data [18] for numerical solutions. However, the uncertainty inherent in inexact methods of solving Einstein's equations, and the importance of obtaining exact solutions wherever possible, has been stressed by Ellis [19]. An exact analytic radiation solution can be obtained, in principle, from an extension of the work reported here, as outlined in section 4. If this can be achieved in practice, even for a restricted spatial geometry, such as that of section 6, the result will serve as a validation test for numerical codes.

Numerical methods are inherently not suited to the investigation of fields in the vicinity of singularities and the curvature singularity revealed by expression (5.9) has not previously been reported. Recent evidence concerning the presence of black holes in galactic nuclei [20] and evidence linking galactic collisions to quasars [21] suggest that the exact solution for the strong and rapidly varying fields in the vicinity of this singularity is likely to lead to valuable insights concerning physically observable processes.

*Acknowledgements*
I would like to thank the following for encouragement and helpful discussions: G.F.R. Ellis and the members of the cosmology group at the University of Cape Town, Sir Herman Bondi, N.T. Bishop, N. Ibragimov, M. MacCallum, R. Penrose, and H. Stephani. I would also like to thank the referees for their comments.

**Appendix A: Symmetry transformations of the hypersurface equations.**
The particular forms of equations (4.1), together with those corresponding to the remaining NP vacuum equations expressed in the coordinates employed in this paper, are determined entirely by the assumptions used in section 2 to construct the coordinate system. These assumptions are not affected by the location from which the null cones, which define the coordinates, are observed. A surface v = const can be marked by a physical light pulse. Provided the metric of the space-time is known, observers positioned at different locations can follow the surfaces v = const as they expand through space by monitoring the reflection of the corresponding light pulses from test objects. Their measurements will be consistent with the NP equations. Under the following conditions the transformations that map the measurements of the observers onto one another will constitute Lie point symmetries of the NP 'hypersurface' equations.

Let the observers all be members of a set S who are located on the axis of symmetry in a regular neighbourhood of O and let them use clocks that have been synchronised at a single event when they all coincided. The axis of symmetry defines an invariant direction common to all the members of the set S. Since $g_{23}$ is zero on the axis of symmetry for every member of the set S in the original coordinates, before applying the transformation (2.2), the same physical light pulse v = const, can be used to identify and label the corresponding hypersurface u = const in the frame of reference of each observer. If, furthermore, each observer uses a boost to match the intervals between light pulses, emitted at intervals dv and received directly from O, to his own clock, then each observer can set the function λ in (2,2) to unity on the axis, so that du = dv for every observer in the set S. Within the single null hypersurface under consideration the different observers are then distinguished only by a single parameter, being their relative separation along the axis.

An observer will, in general, assign the label u = const to a pulse of light emitted at a different time for each angle, such that the intersection of the null surface u = const with a surface of simultaneity, or spacelike slice, in the coordinates of that observer (the 'wave front'), constitutes a sphere on which the coordinate r is the same on the null geodesics corresponding to every angle. This follows from the fact that when $g_{23}$ is zero, the direction $(\partial/\partial\theta)^\mu$ is orthogonal to the direction $(\partial/\partial r)^\mu$. In the coordinates of each observer, the measurements of r, θ, and φ on these surfaces u = const will be related to the values of q, Q, and $\Psi_0$ on the same surfaces by (4.1).

Transformations from the reference frame of one observer in the set S to that of another can be made as follows. The value of r can be deduced from measurements carried out on the wave front v = const in a small neighbourhood of the axis of symmetry. Since $g_{23}$ is zero on the axis of symmetry for every observer in the set S it follows that for a spacelike displacement ds within the hypersurface v = const:

$$\lim_{\theta \to 0} ds^2 = -\frac{1}{2} q^2 d\theta^2 \tag{A1}$$

Since σ = 0 on the axis of symmetry, it follows from (3.6) and (3.2) (setting $\rho^0$ to zero) that

$$\lim_{\theta \to 0} q = q^0 r \tag{A2}$$

where $q^0$ is a function that is independent of r. If one confines the argument to the first quadrant, as is possible without loss of generality, then θ is single-valued and constant on the axis of symmetry. On the axis, $q^0$ can therefore only depend on u. It follows that:

$$r = \frac{\sqrt{2}}{q^0} \lim_{\theta \to 0} \frac{ds}{d\theta} \tag{A3}$$

Points in the hypersurface can be located by specifying s and θ as independent variables, instead of r and θ, with s = θ = 0 on the axis. (There is a coordinate singularity on the axis when using s and θ, analogous to the coordinate singularity at the origin of a spherical coordinate system. However, values of r corresponding to events on the axis in the wave front can be assigned by analytic continuation.) Under a change of basis, in general, $s' = s'(s, \theta)$ and $\theta' = \theta'(s, \theta)$. It follows that under a change of basis within the set S:

$$r' = \frac{\sqrt{2}}{q^0} \lim_{\theta \to 0} \frac{ds'}{d\theta'} = \frac{\sqrt{2}}{q^0} \lim_{\theta \to 0} \frac{\frac{\partial s'}{\partial s} ds + \frac{\partial s'}{\partial \theta} d\theta}{\frac{\partial \theta'}{\partial s} ds + \frac{\partial \theta'}{\partial \theta} d\theta} \tag{A4}$$

If the limits of the partial derivatives in the above expression are independent of r, depending only on the relative positions of the observers, then they are constants and the expression can be re-arranged to read:

$$r' = \frac{\alpha r + \beta}{\gamma r + \delta} \tag{A5}$$

where α, β, γ, and δ are constants. Without loss of generality the condition αδ - γβ = 1 can be imposed. This is then the three-dimensional Lie group SL(2,R).

This group of transformations leaves the NP equations unchanged. However, the function $\Psi_0$ is not determined by the NP equations [5]. In general $\Psi_0$ will be changed by the transformation (A5) so that the transformation, when applied to a solution of the NP equations in the current form, will yield a different solution of the same equations. This is a particular case of a general result, discovered by Geroch [13], that the action of the group SL(2,R) on any exact solution of Einstein's vacuum field equations with a Killing vector produces another exact solution that also possesses a Killing vector. Geroch's result confirms that the partial derivatives in (A4) must be independent of r.

By construction, the map from the manifold defined by the set of points $\Sigma = (r, \theta \,|\, u, \phi = \text{const})$, corresponding to any observer in the set S, to the manifold defined by the set of points $\Sigma' = (r', \theta' \,|\, u', \phi = \text{const})$ corresponding to any other observer in the set S, is a diffeomorphism. (To give the sets Σ and Σ' a manifold structure the singular point O must be excluded). Transformations from Σ to Σ' therefore follow the normal tensor transformation laws. It follows that $\Psi_0$, which is a scalar function, is form-invariant under a transformation from Σ to Σ'.

This result means that both the NP equations and the form of $\Psi_0$ on the initial null hypersurface must remain unchanged under the action of a transformation from the measurements of one observer in the set S to those of another. Since these transformations conform to the axioms for a group of Lie point symmetries, the corresponding transformations of the dependent variables q and Q must be part of a Lie group. The result (4.3) shows that, to meet this requirement, the transformation of r must consist of a linear combination of the elements of the group SL(2,R), and also that the dependent variables q and Q have to transform, uniquely, as determined by the infinitesimal generator $X_1$. This restricts the form of $\Psi_0$ to (4.2) and limits the set of solutions of the NP equations accordingly.

Formal solutions of the NP equations may exist that correspond to functional forms of $\Psi_0$ different from (4.2), but they will not be physically realistic representations of the geometry of the space-time under consideration in this paper. The above analysis has therefore proved the following theorem:

Theorem: *Let M be a simply connected, analytic neighbourhood of a point O, on the axis of symmetry, that generates a timelike curve in an axially symmetric vacuum space-time in which Einstein's field equations are satisfied. Let the space-time possess reflection symmetry with respect to rotation about the axis of symmetry. Let M satisfy the axioms for a space with spinor structure. Consider a null hypersurface u = const constructed according to expression (2.2) from a sequence of null cones generated from O. Then, if r is an affine parameter that measures radial distances from O along the null geodesics that pass through O, the r-dependence of the values of the component $\Psi_0$ of the Weyl spinor on the null hypersurface, u=const, in M is given by the function (4.2) where the $b_i$ are constants and $a_0$ is a function of $\theta$.*

**Appendix B**

The three cases of the solutions of the Sachs equations with $\Psi_0$ in the form (4.2) not mentioned in the text are as follows, putting $b_2 = 1$:

If $b_1^2 - 4b_0 < 0$ but $b_1^2 - 4b_0 + 4|a_0| > 0$, then

$$Q = Q^0 \left(r^2 + b_1 r + b_0\right)^{\frac{1}{2}} \cosh\left\{\left[\frac{b_1^2 - 4(b_0 - a_0)}{4b_0 - b_1^2}\right]^{\frac{1}{2}} \arctan\left[\frac{(4b_0 - b_1^2)^{\frac{1}{2}}}{2r + b_1}\right] + \phi_Q\right\}$$

$$q = q^0 \left(r^2 + b_1 r + b_0\right)^{\frac{1}{2}} \cos\left\{\left[\frac{4(b_0 + a_0) - b_1^2}{4b_0 - b_1^2}\right]^{\frac{1}{2}} \arctan\left[\frac{(4b_0 - b_1^2)^{\frac{1}{2}}}{2r + b_1}\right] + \phi_q\right\}$$

if $a_0 > 0$. If $a_0 < 0$ the forms of q and Q are interchanged.

If the quadratic has two equal roots, i.e. $b_1^2 = 4b_0$ then:

$$Q = Q^0 \left(r + \frac{b_1}{2}\right) \cosh\left(\frac{\sqrt{a_0}}{r + \frac{1}{2}b_1} + \phi_Q\right)$$

$$q = q^0 \left(r + \frac{b_1}{2}\right) \cos\left(\frac{\sqrt{a_0}}{r + \frac{1}{2}b_1} + \phi_q\right)$$

for $a_0 > 0$. If $a_0 < 0$ the forms of q and Q are interchanged and the sign of $a_0$ is changed.

If the quadratic has two unequal roots a and b and $b_1^2 - 4b_0 - 4|a_0| < 0$ then

$$q = q^0 [(r-a)(r-b)]^{\frac{1}{2}} \cos\left\{\left[\frac{a_0}{(a-b)^2} - \frac{1}{4}\right]^{\frac{1}{2}} \ln\left(\frac{r-a}{r-b}\right) + \phi_q\right\}$$

$$Q = Q^0 [(r-a)(r-b)]^{\frac{1}{2}} \cosh\left\{\left[\frac{a_0}{(a-b)^2} + \frac{1}{4}\right]^{\frac{1}{2}} \ln\left(\frac{r-a}{r-b}\right) + \phi_Q\right\}$$

if $a_0 > 0$. If $a_0 < 0$ the forms of q and Q are interchanged and the sign of $a_0$ is changed.

In all cases cosines can be replaced with sines and vice versa. Also, since the equations are linear, a sum of the cosine and sine forms can be used as a solution in each case with the result that the hyperbolic cases can take the form of exponentials.

[1] Bondi, H., 1960 Nature, Lond., 186, 535
[2] Bondi, H., van der Burg, M.G.J. and Metzner, A.W.K. 1962 Proc. Roy. Soc. Lond. A 269, 21
[3] Newman, E. and Penrose, R. 1962 J. Math. Phys. 3, 566 ; errata: ibid. 4, 998
[4] Robinson, I. and Trautman, A. 1960 Phys. Rev. Letters 4, 431
[5] Penrose, R. 1963 Aerospace Research Laboratories 63-65 (P.G. Bergman) reprinted 1980 in Gen. Rel. Grav. 12, 3 , 225
[6] see Penrose, R. and Rindler, W. 1984 "Spinors and Space-time" vol. 1 Cambridge: Cambridge University Press.
[7] see e.g. Wald, Robert M. 1984, *General Relativity* Chicago: University of Chicago Press appendix B3
[8] Sachs, R. 1961 Proc. Roy. Soc. Lond. A 264, 309
[9] Chrusciel, P. T., MacCallum, M.A.H., and Singleton, D.B. 1995 Phil. Trans. R. Soc. Lond. (A ) 350 113
[10] see e.g. Ibragimov, N.H. 1995 "CRC Handbook of Lie group analysis of differential equations" CRC Press
[11] see e.g. Stephani, H. 1989 "Differential equations: Their solution using symmetries" Cambridge: Cambridge University Press p 159
[12] Wessels, E.J.H. 1997 "Second Conference of the Southern African Relativity Society: Participants and Abstracts", Ed. N Bishop, F Bullock and S Maharaj , University of South Africa, Pretoria.
[13] Geroch, R. 1971 J. Math. Phys. 12, 918
[14] Friedrich, H. 1982 Proc. Roy. Soc. Lond. A 381, 361
[15] For a review see Seidel, E. and Suen, W-M. 1994 Int. J. Mod. Phys. C 5, 181.
[16] Matzner, R.A., Seidel, H.E., Stuart, L., Shapiro, L. Smarr, L., Suen, W-M., Teukolsky, S., Winicour, J. "Geometry of a Black Hole Collision" (available on the internet and to be submitted to Science)
[17] Misner, C. 1963 Ann. Phys. (NY) 24, 102
[18] see e.g. Lousto, C.O. and Price, R.H. 1998 Phys. Rev. D 57, 1073
[19] Ellis, G.F.R. 1993 "Exact and inexact solutions of the Einstein field equations". In: "The Renaissance of General Relativity and Cosmology", Ed G Ellis, A Lanza and J Miller. Cambridge University Press, pp 20-39.
[20] For a recent review see Kormendy, J. and Richstone, D. 1995 Annu. Rev. Astron. Astrophys. 33, 581
[21] See e.g. Disney, M.J. et. al. 1995 Nature, 376, 150.